\documentclass[preprint,aps,prc,showpacs,nofootinbib]{revtex4}
\usepackage{amsmath}
\usepackage{amssymb}
\usepackage{graphics}
\usepackage{epsfig}

\usepackage{subfigure}

\newcommand\nn{\nonumber}
\newcommand\ba{\begin{eqnarray}}
\newcommand\ea{\end{eqnarray}}
\newcommand\be{\begin{equation}}
\newcommand\ee{\end{equation}}

\begin{document}

\title{Polarization observables in lepton-deuteron elastic scattering including the lepton mass}

\author{G. I. Gakh} 
\affiliation{\it National Science Centre, Kharkov Institute of Physics and Technology, \\ 61108 Akademicheskaya 1, Kharkov, and V. N. Karazin Kharkov National University, Dept. of
Physics and Technology, 31 Kurchatov, 61108, Kharkov, Ukraine}
\author{A. G. Gakh}
\affiliation{\it V. N. Karazin Kharkov National University, Dept. of
Physics and Technology, 31 Kurchatov, 61108, Kharkov, Ukraine}

\author{E. Tomasi--Gustafsson}
\email{etomasi@cea.fr}
\affiliation{\it DSM/IRFU/SPhN, CEA/Saclay, 91191 Gif-sur-Yvette, and Universit\'e Paris-Sud, CNRS/IN2P3, Institut de Physique Nucl\'eaire, UMR 8608, 91405 Orsay, France}


\begin{abstract}
Expressions for the
unpolarized differential cross section and for various polarization
observables in the lepton-deuteron elastic scattering, $\ell+D\to
\ell+D$, $\ell=e$, $\mu$, $\tau$, have been obtained in one-photon-exchange approximation, taking into account the lepton mass.  Polarization effects have
been investigated for the case of a polarized lepton beam and
polarized deuteron target which can have vector or tensor
polarization.  Numerical
estimations of the lepton mass effects have been done for the
unpolarized differential cross section and for some polarization
observables and applied to the case of low energy muon deuteron elastic scattering.
\end{abstract}

\maketitle
\section{Introduction}
\label{eq:Introduction}

The structure of hadrons and nuclei is traditionally studied through
elastic and inelastic electron-hadron (nuclei) scattering assuming
one-photon-exchange mechanism. A review of the results obtained by
the measurements of the unpolarized cross section and polarization
observables in the elastic electron-nucleon scattering can be found
in Ref. \cite{PPV}. Recent review of the nucleon form factors in the
time-like region can be found in Ref. \cite{Denig:2012by}. 
A review of the deuteron electromagnetic structure is given in 
Ref. \cite{GOGG}.

Recently, results from the measurement of the
proton charge radius were obtained in an experiment performed at PSI
(Switzerland) \cite{P10} from the Lamb shift in
muonic hydrogen (CREMA collaboration). The obtained value is significantly different from earlier measurements based on electronic hydrogen
spectroscopy and elastic electron-proton scattering and it is  smaller by  7$\sigma $ than the 2010 CODATA official value \cite{MTN12}. Various explanations of this difference were proposed.

Some authors suggested the possible existence of new particles that interact with muons and hadrons but not with
electrons. Adjusting the couplings of these particles one can, in
principle, obtain an additional energy shift in the muonic hydrogen.
This may lead to the agreement between the measurement of the proton
charge radius in the muonic and electronic experiments. Thus, for
example, the existence of new particles with scalar and pseudoscalar
(or vector and axial) couplings were proposed in Ref. \cite{Carlson:2012pc}.
The couplings are constrained by the existing data on the Lamb
shift, muon magnetic moment and kaon decay rate. New vector and
scalar particles at the 100 MeV scale were proposed in Ref. \cite{BKP}. The important
consequence would be an enhancement by several orders of magnitude
of the parity violating asymmetries in the scattering of low-energy
muons on nuclei.

On the other hand, the authors of the Ref. \cite{Lorenz:2012tm} have analyzed
the recent electron-proton scattering data obtained at Mainz
\cite{B10} (the cross sections were measured with statistical errors
below $0.2\%$). Using a dispersive approach they obtained a small
value for the proton charge radius which is consistent with the
recent result obtained in the experiment with muonic hydrogen. In
Ref. \cite{Paz:2011qr} it was shown that previous extractions of the proton
charge radius from the electron-proton scattering data may have
underestimated the errors. 

In electron-proton elastic scattering experiments, the radius is related to the slope of the charge form factors in the limit of the transferred momentum squared $Q^2\to 0$. In Ref. 
\cite{Gakh:2013pda} it was suggested 
that the error related to the extrapolation $Q^2\to 0$ could be reduced by measuring this process in inverse kinematics.

Note that about 40 years ago there were tests of the muon-electron
universality in the processes of the elastic and deep inelastic
electron (muon) scattering. Measurements of the muon-proton elastic
cross section in the range $0.15\le Q^2\le 0.85$ GeV$^2$  were
compared with similar electron-proton data \cite{C69}. It was found
an apparent disagreement between muon and electron experiments which
can possibly be accounted for by a combination of systematic
normalization errors \cite{C69}. The data were obtained at rather
high values of $Q^2$ in order to extract the proton charge radius.
In Ref. \cite{K74}, the muon-proton elastic
scattering was measured in the range $0.6\le Q^2\le 3.2$ GeV$^2$. A possible difference from muon-electron universality was found, but
the statistical accuracy of this observation was not compelling. The
muon-proton deep inelastic scattering was measured in the range
$0.4\le Q^2\le 3.6$ GeV$^2$ \cite{E74}. The data were consistent
with muon-electron universality. Two-photon-exchange effects were
investigated in the muon-proton elastic scattering \cite{CA69}. The validity of the one-photon-exchange approximation was confirmed for $Q^2$ up to 0.85 GeV$^2$ and incident muon energies up to 17 GeV.

The fact that the proton charge radius was not measured in the
process of the elastic muon-proton scattering led to the proposal of
the MUon proton Scattering Experiment (MUSE) at the Paul Scherrer
Institut (Zurich) \cite{Gilman:2013eiv}. This experiment plans 
a simultaneous measurement of the elastic $\mu^-p$ and $e^-p$ scattering  as well as $\mu^+ p$ and $e^+p$ and it will establish the consistency or the 
difference of the muon-proton and electron-proton interaction with good precision in the considered kinematics \cite{Gilman:2013eiv}. Three values of the muon beam momenta which are comparable with the muon mass: 115 MeV, 153 MeV, and 210
MeV, were chosen.  However, in case of low
energy and large lepton mass, the terms proportional to the lepton
mass become important and the mass should be taken explicitly into
account in the calculation of the kinematical variables and of the experimental observables. The expressions of the kinematical relations and of the
polarized and unpolarized observables are different from those
currently used. In Ref. \cite{P87} the effect of the lepton mass was
discussed for muon-proton elastic scattering for the unpolarized
cross section and the double spin asymmetry, where the lepton beam
and the target are polarized.

The MUSE experiment at PSI will also determine the radii of
light nuclei, through muon elastic scattering. Of particular interest is a
measurement on deuterium. The issue of taking into account finite lepton mass effects is also relevant for the case of the elastic
muon-deuteron scattering. 

In this paper we calculate the expressions for the unpolarized
differential cross section and polarization observables, taking into
account the lepton mass, for elastic lepton-deuteron scattering. We
calculate the asymmetries due to the tensor polarization of the
deuteron target and the spin correlation coefficients due to the
lepton beam polarization and vector polarization of the deuteron
target. Explicit formulas are given in two coordinate systems which are relevant for the experiment: in
the first one the $z$ axis is directed along the lepton beam momentum and
in the second one - along the virtual photon momentum (or along the
transferred momentum).

\section{Formalism}
Let us consider the reaction:
\be
\ell(k_1)+D(p_1) \to \ell(k_2) +D(p_2),~\ell=e,~\mu,~\tau, 
\label{eq:1}
\ee
where the momenta of the particles are written in the parenthesis.
In the laboratory system, where we perform our analysis,  the deuteron
(lepton) four momenta in the initial and final states are respectively
$p_1$ and $p_2$ ($k_1$ and $k_2$) with components:
\be\label{eq:2}
p_1=(M,0),~ p_2=(E_2, \vec p_2),~ k_1=(\varepsilon_1, \vec k_1),~
k_2=(\varepsilon_2, \vec k_2),
\ee
where $M$ is the deuteron mass.

The matrix element of the reaction (\ref{eq:1}) can be written as follows in
the one-photon-exchange approximation
\be\label{eq:3}
{\cal M}=\frac{e^2}{Q^2}j_{\mu}J_{\mu},~j_{\mu}=\bar u(k_2)\gamma_{\mu }u(k_1).
\ee
Using the requirements of the Lorentz invariance, current
conservation, parity and time-reversal invariance of the hadron
electromagnetic interaction, the general form of the electromagnetic
current for the spin-one deuteron is completely described by three
form factors and it can be written, following Ref. \cite{GKM}, as
\ba
J_{\mu}&=&(p_1+p_2)_{\mu}\Big[-G_1(Q^2)U_1\cdot U_2^*+ 
 \frac{G_3(Q^2)}{M^2} (U_1\cdot q U_2^*\cdot
q-\frac{q^2}{2}U_1\cdot U_2^*)\Big] 
\nn\\ 
&&
+G_2(Q^2)(U_{1\mu}U_2^*\cdot q-U_{2\mu}^*U_1\cdot q),
\label{eq:4}
\ea
where $q=k_1-k_2=p_2-p_1,$ $ Q^2=-q^2 $, $U_{1\mu}$ and $U_{2\mu}$
are the polarization four-vectors for the initial and final deuteron
states. The functions $G_i(Q^2)$, $i=1, 2, 3$, are the deuteron
electromagnetic form factors, which are real functions in the region of the
space-like momentum transfer and depend only on the virtual photon
four-momentum squared. 
These form factors are related to the standard deuteron form
factors: $G_C$ (the charge monopole), $G_M$ (the magnetic dipole)
and $G_Q$ (the charge quadrupole) by 
\be\label{eq:5}
G_M=-G_2, \ G_Q=G_1+G_2+2G_3, \ G_C=\frac{2}{3}\tau (G_2-G_3)+
(1+\frac{2}{3}\tau )G_1,
\ee
with $\tau =Q^2/(4M^2)$. The standard form factors have the following normalization:
$$ G_C(0)=1\,, \ G_M(0)=(M/m_n)\mu_d\,, \
G_Q(0)=M^2Q_d\,,$$ where $m_n$ is the nucleon mass, $\mu_d(Q_d)$ is
deuteron magnetic (quadrupole) moment and their values are:
$\mu_d=0.857 $ \cite{MT}, $Q_d=0.2859 fm^2$ \cite{ERC}.

The differential cross section can be written in terms of the matrix element modulus squared as
\be\label{eq:6}
d\sigma=\frac{(2\pi )^4}{4I}| M|^2\frac{d{\vec k}_2d{\vec
p}_2}{(2\pi )^64\varepsilon_2E_2}\delta^{(4)}(k_1+p_1-k_2-p_2),
\ee
where $I^2=(k_1\cdot p_1)^2-m^2M^2$ and $m$ is the lepton mass.

Writing the matrix element in the form 
${\cal M}=(e^2/Q^2){\overline {\cal M}}$ one obtains the following expression for the
differential cross section of the reaction (\ref{eq:1}) in the laboratory
system for the case when the scattered lepton is detected in the final
state
\be\label{eq:7}
\frac{d\sigma}{d\Omega}=\frac{\alpha ^2}{4M}\frac{{\vec
k}_2^2}{d|{\vec k}_1|}\frac{|{\it \overline M}|^2}{Q^4},
\ee
where $d=(M+\varepsilon_1)|{\vec k}_2|-\varepsilon_2|{\vec
k}_1|\cos\theta  $, $\theta $ is the lepton scattering angle (angle
between the initial and final lepton momenta). The scattered lepton
energy has the following form in term of the lepton scattering angle
\be\label{eq:8}
\varepsilon_2=\frac{(\varepsilon_1+M)(M\varepsilon_1 +
m^2)+\vec{k}_1^2\cos\theta\sqrt{ M^2-m^2\sin^2\theta}}
{(\varepsilon_1+M)^2-\vec{k}_1^2\cos^2\theta }.
\ee
In the limit of zero lepton mass this expression gives the well known
relation between the energy and angle of the scattered lepton:
$$\varepsilon_2=\varepsilon_1\left [1+2(\varepsilon_1/M)\sin^2\frac{\theta}{2}\right ]^{-1}.$$
The differential cross section for the case when the recoil deuteron
is detected in the final state can be written as
\be\label{eq:9}
\frac{d\sigma}{d\Omega_D}=\frac{\alpha ^2}{4M}\frac{{\vec p}_2
^2}{\bar d|{\vec k}_1|}\frac{|{\overline {\cal M}}|^2}{Q^4},
\ee
where $\bar d=(M+\varepsilon_1)|{\vec p}_2|-E_2|{\vec
k}_1|\cos\theta_D $, $\theta_D $ is the angle between the momenta of
the lepton beam and recoil deuteron. Using the relation
\be\label{eq:10}
dQ^2=|{\vec k}_1||{\vec
p}_2|\frac{1}{\pi}\frac{E_2+M}{\varepsilon_1+M}d\Omega_D
\ee
we obtain the following expression for the differential cross section over the
$Q^2$ variable
\be\label{eq:11}
\frac{d\sigma}{dQ^2}=\frac{\pi\alpha ^2}{4M}\frac{{|\vec p}_2|}{\bar
d{\vec k}_1^2}\frac{\varepsilon_1+M}{E_2+M}\frac{|{\overline{\cal M}}|^2}{Q^4}.
\ee

The square of the reduced matrix element can be written as
\be\label{eq:12}
|{\overline {\cal M}}|^2 =L_{\mu\nu }H_{\mu\nu },
\ee
where the leptonic $L_{\mu\nu }$ and hadronic $H_{\mu\nu }$ tensors
are defined as follows
\be\label{eq:13}
L_{\mu\nu }=j_{\mu}j_{\nu }^*, \ \ H_{\mu\nu }=J_{\mu}J_{\nu }^*.
\ee
If the initial and scattered leptons are unpolarized then in this case
the leptonic tensor is
\be\label{eq:14}
L_{\mu\nu }(0)=2q^2g_{\mu\nu}+4(k_{1\mu}k_{2\nu}+k_{2\mu}k_{1\nu}).
\ee
In the case of polarized lepton beam the spin-dependent part of
the leptonic tensor can be written as
\be\label{eq:15}
L_{\mu\nu }(s)=2im<\mu\nu qs_l>,
\ee
where $<\mu\nu ab>=\varepsilon_{\mu\nu\rho\sigma
}a_{\rho}b_{\sigma}$ and $s_{l\mu}$ is the lepton polarization
4-vector which satisfies the conditions $s_l^2=-1, k_1\cdot s_l=0$.

For an arbitrary polarization state of the initial and
recoil deuterons, we may write the electromagnetic current in the following form
$$J_{\mu}=J_{\mu\alpha\beta}U_{1\alpha}U_{2\beta}^*,$$
and the hadronic tensor $H_{\mu\nu}$ becomes 
\be\label{eq:16}
H_{\mu\nu}=J_{\mu\alpha\beta}J_{\nu\sigma\gamma}^*\rho_{\alpha\sigma}^i
\rho_{\gamma\beta}^f,
\ee
where $\rho_{\alpha\sigma}^i \ (\rho_{\gamma\beta}^f)$ is the
spin-density matrix of the initial (final) deuteron.

As we consider the case of a polarized deuteron target and
unpolarized recoil deuteron, the hadronic tensor $H_{\mu\nu}$ can be
expanded according to the polarization state of the initial deuteron: 
\be\label{eq:17}
H_{\mu\nu}=H_{\mu\nu}(0)+H_{\mu\nu}(V)+H_{\mu\nu}(T),
\ee
where the spin-independent tensor $H_{\mu\nu}(0)$ corresponds to an
unpolarized initial deuteron and the spin-dependent tensor
$H_{\mu\nu}(V) \ (H_{\mu\nu}(T))$ describes the case where the
deuteron target has a vector (tensor) polarization.

In the general case, the initial deuteron polarization
state is described by the spin-density matrix. The general
expression for the deuteron spin-density matrix in the coordinate
representation is \cite{Schildknecht:1965}
\be\label{eq:18}
\rho_{\alpha\beta}^i=-\frac{1}{3} \left (g_{\alpha\beta}-\frac{p_{1\alpha}
p_{1\beta}}{M^2} \right ) +\frac{i}{2M}<\alpha\beta
sp_1>+Q_{\alpha\beta} \ ,
\ee
where $s_{\mu}$ is the polarization four-vector describing the
vector polarization of the deuteron target $(p_1\cdot s=0,~ 
s^2=-1)$ and $Q_{\mu\nu}$ is the tensor which describes the  quadrupole polarization of the initial deuteron and satisfies the
following conditions: $Q_{\mu\nu}=Q_{\nu\mu}$, $Q_{\mu\mu}=0$, $
p_{1\mu}Q_{\mu\nu}=0$. In the laboratory system (the initial
deuteron rest frame) all time components of the tensor $Q_{\mu\nu}$
are zero and the tensor polarization of the deuteron target is
described by five independent space components 
$$Q_{ij}=Q_{ji}, \ Q_{ii}=0, \ i,j=x,y,z.$$
If the polarization of the recoil deuteron is not
measured, the deuteron spin-density matrix can be written as
\be\label{eq:19}
\rho_{\alpha\beta}^f=-\Big(g_{\alpha\beta}-\frac{p_{2\alpha}
p_{2\beta}}{M^2}\Big).
\ee
The relation between elements of the
deuteron spin-density matrix in the helicity and spherical tensor
representations as well as in the coordinate representation is given in the Appendix. The relations between the polarization parameters $s_i$ and $Q_{ij}$ and the population numbers $n_+, \ n_-$ and $n_0$
describing the polarized deuteron target, which is often used in
spin experiments, are also given.

\section{Unpolarized differential cross section}

Let us consider the elastic scattering of unpolarized lepton beam by unpolarized deuteron target. The hadronic tensor
$H_{\mu\nu}(0)$ can be written as
\be\label{eq:20}
H_{\mu\nu }(0)=H_1(Q^2) \tilde{g}_{\mu\nu }+H_2(Q^2)\tilde p_{1\mu
}\tilde p_{1\nu },~\tilde{g}_{\mu\nu}=g_{\mu\nu}-\frac{q_{\mu}q_{\nu}}{q^2}\ , \ \
\tilde{p}_{1\mu}=p_{1\mu}-\frac{p_1\cdot q}{q^2}q_{\mu} \ . 
\ee 
The real structure functions $H_{1,2}(Q^2)$ are expressed in terms of
the deuteron electromagnetic form factors as
\be\label{eq:21}
H_1(Q^2)=\frac{2}{3}Q^2 (1+\tau )G_M^2, \ \
H_2(Q^2)=4M^2(G_C^2+\frac{2}{3}\tau G_M^2+\frac{8}{9}\tau ^2G_Q^2).
\ee
The contraction of the spin-independent leptonic $L_{\mu\nu}(0)$ and
hadronic $H_{\mu\nu}(0)$ tensors gives
\be\label{eq:22}
S(0)=-4(q^2+2m^2)H_1(Q^2)+2[(1+\tau )q^2+\frac{4}{M^2}(k_1\cdot
\tilde p_1)^2]H_2(Q^2),
\ee
where the averaging over the spin of the initial deuteron is
included in the  structure functions $H_{1,2}(Q^2)$.

Substituting this expression into Eq. (\ref{eq:7}) and averaging over
the spin of the initial lepton, we obtain the expression for the
unpolarized differential cross section of the reaction (\ref{eq:1}) in the
laboratory system, taking into account the lepton mass, in the form
\be\label{eq:23}
\frac{d\sigma_{un}}{d\Omega } = \sigma_0D,
\ee
where $\sigma_0 $ is the cross section for the scattering of lepton
on the point spin 1 particle. It is a generalization of the Mott
cross section (with a recoil factor) to the case when the lepton
mass is not neglected
\be\label{eq:24}
\sigma_0=4\frac{\alpha^2}{q^4}\frac{M}{d}\frac{{\vec k}_2^2}{|{\vec
k}_1|}[\varepsilon_1^2-M(M+2\varepsilon_1)\tau ].
\ee
Note that in the limit $m=0$ this expression reduces to the Mott
cross section
\be\label{eq:25}
\sigma_0(m=0)=\sigma_{Mott}=\frac{\alpha^2\cos^2\frac{\theta}{2}}
{4\varepsilon_1^2\sin^4\frac{\theta}{2}}\left (1+
2\frac{\varepsilon_1}{M}\sin^2\frac{\theta}{2}\right )^{-1},
\ee
where $\theta $ is the lepton scattering angle (between the momenta of
the initial and final leptons).

The quantity $D$, which contains the information about the structure
of the deuteron, has a form
\be\label{eq:26}
D=A(Q^2)+f(Q^2, \varepsilon_1, m)B(Q^2),
\ee
where the standard structure functions
$A(Q^2)$ and $B(Q^2)$ which describe unpolarized differential cross
section of the reaction (\ref{eq:1}) in the zero lepton mass approximation are explicitly singled out: 
\be\label{eq:27}
A(Q^2)=G_C^2(Q^2)+\frac{8}{9}\tau ^2G_Q^2(Q^2)+\frac{2}{3}\tau
G_M^2(Q^2), \ \ B(Q^2)=\frac{4}{3}\tau (1+\tau )G_M^2(Q^2).
\ee
The function $f(Q^2, \varepsilon_1, m)$ has a form
\be\label{eq:28}
f(Q^2, \varepsilon_1,
m)=(Q^2-2m^2)\left [4\varepsilon_1^2-Q^2\left (1+2\frac{\varepsilon_1}{M}\right )\right ]^{-1}.
\ee
In the limit of zero lepton mass this function reduces to
$$f(Q^2, \varepsilon_1,m=0)=\tan^2\frac{\theta}{2}.$$ 
Thus, in this approximation we obtain
the standard expression for the unpolarized differential cross
section of the reaction (\ref{eq:1})
\be\label{eq:29}
\frac{d\sigma_{un}}{d\Omega
}=\sigma_{Mott}\Big\{A(Q^2)+B(Q^2)\tan^2(\frac{\theta} {2})\Big\}.
\ee
In the standard approach (zero lepton mass) the measurement of the
unpolarized differential cross section at various values of the lepton
scattering angle and the same value of $Q^2$ allows to determine the
structure functions $A(Q^2)$ and $B(Q^2)$. Therefore, it is possible
to determine the magnetic form factor $G_M(Q^2)$ and the following
combination of the form factors $G_C^2(Q^2)+8\tau^2G_Q^2(Q^2)/9$. 

The determination of this quantity in the non-zero lepton mass approximation, requires the measurement of the unpolarized differential cross section at different
values of the lepton beam energy and at the same value of $Q^2$. 

The separation of the charge $G_C$ and quadrupole $G_Q$ form factors
requires polarization measurements.

\section{Vector polarized deuteron target }

The calculation of polarization observables requires 
to choose a coordinate frame. Let us define the following coordinate frame in
the Lab system: the $z$-axis is directed along the lepton beam
momentum $\vec{k}_1$, the $y$-axis is directed along the vector
$\vec{k}_1\times \vec{k}_2$, and the $x$-axis is chosen in order to
form a left handed coordinate system. Therefore the reaction plane is the
$xz$-plane.

In this section we consider the T-even polarization observables,
which depend on the spin correlation $\vec{s}\cdot\vec{s}_l$ which
determine the scattering of the polarized lepton beam (of spin $s_l$) by vector
polarized deuteron target (of spin $s$).

In the considered case the spin-dependent tensor $H_{\mu\nu}(V)$,
that describes the vector polarized initial deuteron and unpolarized
final deuteron, can be written as
\ba
H_{\mu\nu}(V)&=&\frac{i}{M}S_1(Q^2)<\mu\nu sq>+
\nn\\
&&
\frac{i}{M^3}S_2(Q^2)\left [\tilde{p}_{1\mu} <\nu
sqp_1>-\tilde{p}_{1\nu}<\mu sqp_1>\right ]+\nn\\
&&
\frac{1}{M^3}S_3(Q^2)[\tilde{p}_{1\mu} <\nu
sqp_1>+\tilde{p}_{1\nu}<\mu sqp_1>], 
\label{eq:30}
\ea
where the three real structure functions $S_i(Q^2), i=1-3$, can be expressed in terms of the
deuteron electromagnetic form factors as: 
\be\label{eq:31}
S_1(Q^2)=M^2(1+\tau )G_M^2, \ S_2(Q^2)=M^2[G_M^2-2(G_C+\frac{\tau
}{3}G_Q)G_M], \ \ S_3(Q^2)=0.
\ee

The third structure function $S_3(Q^2)$ vanishes since the deuteron form
factors are real functions for elastic scattering (in the space-like region of 
momentum transfer). In the time-like region (for annihilation processes, for example, $e^-+e^+\to
D+\bar D $), where the form factors are complex functions, the
structure function $S_3(Q^2)$ is not zero and it is determined by
the imaginary part of the form factors, namely: $S_3(Q^2)=
2M^2Im(G_C-\tau /3G_Q)G_M^*$.

The differential cross section of the reaction (\ref{eq:1}) describing the
scattering of polarized lepton beam on vector polarized deuteron
target can be written as (referring only to the spin-dependent part
of the cross section which is determined by the spin correlation
coefficients)
\be\label{eq:32}
\frac{d\sigma(s,
s_l)}{d\Omega}=\frac{d\sigma_{un}}{d\Omega}(1+C_{xx}\xi_x\xi_{lx}+
C_{yy}\xi_y\xi_{ly}+C_{zz}\xi_z\xi_{lz}+C_{xz}\xi_x\xi_{lz}+C_{zx}\xi_z\xi_{lx}),
\ee
where the vector $\vec{\xi}_l (\vec{\xi})$ is the unit polarization
vector in the rest frame of the lepton beam (deuteron target). The spin correlation coefficients $C_{ij}$ have the
following form in terms of the deuteron electromagnetic form factors
\ba
DC_{xx}&=&\frac{m}{M}\frac{G_M}{z}\left [\tau \vec{k}_2^2\sin^2\theta
G_M-2\left (|\vec{k}_1|-|\vec{k}_2|\cos\theta \right )^2\left (G_C+\frac{\tau}{3}G_Q\right )\right ],
\nn\\
DC_{yy}&=&2\frac{m}{M}\frac{q^2}{z}(1+\tau
)\left (G_C+\frac{\tau}{3}G_Q\right )G_M, 
\nn\\
DC_{zx}&=&-\frac{m}{M}\frac{|\vec{k}_2|}{z}\sin\theta (|\vec{k}_1|-
|\vec{k}_2|\cos\theta )\left (\tau G_M+2G_C+\frac{2}{3}\tau G_Q\right )G_M,
\label{eq:33}\\
DC_{xz}&=&-\frac{|\vec{k}_2|}{zM}\sin\theta G_M
\left [\varepsilon_1\left (|\vec{k}_1|-|\vec{k}_2|\cos\theta \right )
\left (\tau G_M+2G_C+\frac{2}{3}\tau G_Q\right )
\right .
\nn\\ 
&&
\left . -2M|\vec{k}_1|\tau (1+\tau )G_M\right ], \nn\\
DC_{zz}&=&\frac{G_M}{zM}
\Bigl \{\tau G_M(|\vec{k}_1|-|\vec{k}_2|\cos\theta )
\left [\varepsilon_1(|\vec{k}_1|-|\vec{k}_2|\cos\theta )-2M|\vec{k}_1|
(1+\tau )\right ]
\nn\\ 
&&
 -2\varepsilon_1\vec{k}_2^2\sin^2\theta
\left (G_C+\frac{\tau}{3}G_Q\right )\Bigr \}, 
\nn
\ea
where $z=4[\varepsilon_1^2-\tau
(M^2+2M\varepsilon_1)]$. Note that the spin correlation coefficients
$C_{xx}, C_{yy}$ and $C_{zx}$ correspond to the transverse
(relative to the lepton beam momentum) components of the spin vector
$\vec{\xi}_l$ and therefore they are proportional to the lepton mass.
The spin correlation coefficients $C_{xz}$ and $C_{zz}$ describe the
scattering of the longitudinally polarized lepton beam and they are
not suppressed by the factor $m/M$.

In the limit of zero lepton mass we have the following expressions for
the spin correlation coefficients
\ba
\bar DC_{xz}^{(0)}&=&\frac{1}{2}\frac{\tau}{\varepsilon_1}\tan\frac{\theta}{2}
\left [(\varepsilon_1+\varepsilon_2)G_M-4(M+\varepsilon_1)\left (G_C+\frac{\tau}{3}G_Q\right )\right ]G_M,
\nn\\
\bar DC_{zz}^{(0)}&=&-2\tau G_M\frac{M}{\varepsilon_1}\left [G_C+\frac{\tau}{3}G_Q+
\frac{\varepsilon_2}{2M^2}(M+\varepsilon_1)\left (1+\frac{\varepsilon_1}{M}
\sin^2\frac{\theta}{2}\right )\tan^2\frac{\theta}{2}G_M\right ],
\nn\\
C_{xx}^{(0)}&=&C_{yy}^{(0)}=C_{zx}^{(0)}=0, \ \bar
D=A(Q^2)+B(Q^2)\tan^2\frac{\theta}{2}. 
\label{eq:34}
\ea 
The expressions of these coefficients coincide
with the results obtained in Ref. \cite{GKM}.

Another coordinate system is also used in the description of the
elastic lepton-deuteron scattering: the $Z$-axis is directed along
the virtual photon momentum (transferred momentum) $\vec{q}$, the
$Y$-axis is directed along the vector $\vec{k}_1\times \vec{k}_2$
and it coincides with the $y$ axis, and the $X$-axis is chosen in
order to form a left handed coordinate system. These coordinate
systems are connected by a rotation in the reaction scattering
plane, which it is determined by the angle $\psi $ between the
direction of the lepton beam and the virtual photon momentum:
\be\label{eq:35}
\cos\psi
=\frac{M+\varepsilon_1}{|\vec{k}_1|}\sqrt{\frac{\tau}{1+\tau}}, \
\sin\psi
=-\frac{1}{|\vec{k}_1|}\frac{1}{\sqrt{1+\tau}}\sqrt{\varepsilon_1\varepsilon_2-\tau
M^2-m^2(1+\tau )}.
\ee

Thus the spin correlation coefficients in the new coordinate system
can be related to the ones in the previously considered coordinate system by the
following relations
$$C_{Zz}=\cos\psi C_{zz}+\sin\psi C_{xz}, \ \ C_{Xz}=-\sin\psi C_{zz}+\cos\psi
C_{xz}, $$
\be\label{eq:36}
C_{Zx}=\cos\psi C_{zx}+\sin\psi C_{xx}, \ \ C_{Xx}=-\sin\psi
C_{zx}+\cos\psi C_{xx}.
\ee
The spin correlation coefficient $C_{yy}$ is the same in both
coordinate systems. We do not transform the spin components of the
lepton beam in order not to mix the transverse and longitudinal
components (relative to the lepton beam momentum) since only the last
one leads to the spin correlation coefficients which are not
proportional to the lepton mass. So, in this coordinate system the
$z$-component of the spin of the lepton beam corresponds to the
longitudinal polarization and the $x$-component - to the transverse
polarization which belongs to the reaction scattering plane.

After performing the rotation we obtain the following expressions
for the spin correlation coefficients in the new coordinate system
\ba
DC_{Zz}&=&\frac{1}{Mz|\vec{k}_1|}\frac{G_M}{\sqrt{\tau (1+\tau
)}}\Bigl \{ 2M|\vec{k}_1|\tau (1+\tau )\Bigl [\tau
(M+\varepsilon_1)\left (|\vec{k}_2|\cos\theta
-|\vec{k}_1|\right )-
\nn\\
&&
r|\vec{k}_2|\sin\theta \Bigr ]G_M
+\varepsilon_1\left (|\vec{k}_2|\cos\theta -|\vec{k}_1|\right)\Bigl[\tau^2(M+\varepsilon_1)
\left (|\vec{k}_2|\cos\theta -|\vec{k}_1|\right )-
\nn\\
&&
r|\vec{k}_2|\sin\theta \Bigr ]G_M
-2\varepsilon_1|\vec{k}_2|\sin\theta \Bigl [\tau
(M+\varepsilon_1)|\vec{k}_2|\sin\theta+
\nn \\
&&
r\left(|\vec{k}_2|\cos\theta
-|\vec{k}_1|\right)\Bigr ]\left (G_C+\frac{\tau}{3}G_Q\right )\Bigr \},
\label{eq:37}\\
DC_{Xz}&=&\frac{1}{Mz|\vec{k}_1|}\frac{G_M}{\sqrt{\tau (1+\tau
)}}\Bigl\{2\varepsilon_1|\vec{k}_2|\sin\theta \Bigl [\tau
\left (M+\varepsilon_1\right)\left (|\vec{k}_2|\cos\theta
-|\vec{k}_1|\right)
\nn\\
&&
-r|\vec{k}_2|\sin\theta \Bigr ]\left (G_C+\frac{\tau}{3}G_Q)\right)
+\tau \Bigl [\tau
(M+\varepsilon_1)|\vec{k}_2|\sin\theta 
\nn\\
&&
+r\left (|\vec{k}_2|\cos\theta
-|\vec{k}_1|\right) \Bigr ]\left [\varepsilon_1\left (|\vec{k}_2|\cos\theta
-|\vec{k}_1|\right)+2M|\vec{k}_1|(1+\tau )\right ]G_M\Bigr \},
\nn\\
DC_{Zx}&=&-\frac{m}{M}\frac{1}{z|\vec{k}_1|}\frac{G_M}{\sqrt{\tau (1+\tau
)}}\Bigl \{\tau |\vec{k}_2|\sin\theta \Bigl [r|\vec{k}_2|\sin\theta 
-\tau
(M+\varepsilon_1)
\nn\\
&&
\left (|\vec{k}_2|\cos\theta -|\vec{k}_1|\right )\Bigr ]G_M+ 
2\left (|\vec{k}_1|-|\vec{k}_2|\cos\theta \right)\Bigl [\tau (M+\varepsilon_1)|\vec{k}_2|\sin\theta 
\nn\\
&&
-r\left (|\vec{k}_1|-|\vec{k}_2|\cos\theta \right)\Bigr ]\left (G_C+\frac{\tau}{3}G_Q\right )\Bigr \}, 
\nn\\
DC_{Xx}&=&\frac{m}{M}\frac{1}{z|\vec{k}_1|}\frac{G_M}{\sqrt{\tau (1+\tau
)}}\Bigr \{\tau |\vec{k}_2|\sin\theta \Bigr [\tau
|\vec{k}_2|(M+\varepsilon_1)\sin\theta 
\nn\\
&&
+r\left (|\vec{k}_2|\cos\theta-|\vec{k}_1|\right )\Bigr ]G_M
-2\left (|\vec{k}_1|-|\vec{k}_2|\cos\theta \right )\Bigr[\tau (M+\varepsilon_1)
\nn\\
&&
\left(|\vec{k}_1|-|\vec{k}_2|\cos\theta \right )
+r|\vec{k}_2|\sin\theta \Bigl ]\left (G_C+\frac{\tau}{3}G_Q\right )\Bigr \}, \nn\\
r&=&\sqrt{\tau [\varepsilon_1\varepsilon_2-\tau M^2-m^2(1+\tau )]}.
\nn
\ea
In the limit of zero lepton mass there are only two nonzero spin
correlation coefficients corresponding to the longitudinal
polarization of the lepton beam and they have following form
\ba\label{eq:38}
\bar DC^{(0)}_{Zz}&=&-\tau \sqrt{(1+\tau )\left (1+\tau
\sin^2\frac{\theta}{2}\right )}\tan\frac{\theta}{2}\sec\frac{\theta}{2}G^2_M,
\nn\\
\bar DC^{(0)}_{Xz}&=&-2\sqrt{\tau (1+\tau )}\tan\frac{\theta}{2}
G_M\left (G_C+\frac{\tau}{3}G_Q\right ). 
\ea
Along with the transformation of the spin correlation coefficients
it is necessary to transform the vector which describes the vector
polarization of the deuteron target. The new components of this
vector are related to $s_{z}$, and $s_{x}$, Eq. (\ref{eq:18}), as follows
\be\label{eq:38a}
s_{I}=V_{Ii}(\psi)s_{i}, \ \
V(\psi)=\left(\begin{array}{cc}\cos{\psi}&\sin{\psi}\\
-\sin{\psi}&\cos{\psi}\end{array} \right)\,,
\ee
where $I=Z, X$ and $i=z, x.$
\section{Tensor polarized deuteron target}

In the case of tensor-polarized deuteron target, the general
structure of the spin-dependent tensor $H_{\mu\nu}(T)$ can be
written in terms of five structure functions as follows
\ba\label{eq:39}
H_{\mu\nu}(T)&=&V_1(Q^2)\bar Q\tilde{g}_{\mu\nu}+
V_2(Q^2)\frac{\bar Q}{M^2} \tilde{p}_{1\mu}\tilde{p}_{1\nu}+
V_3(Q^2)(\tilde{p}_{1\mu}\widetilde{Q}_{
\nu}+\tilde{p}_{1\nu}\widetilde{Q}_{\mu})+
\nn\\
&&+V_4(Q^2)\widetilde{Q}_{\mu\nu}+
iV_5(Q^2)(\tilde{p}_{1\mu}\widetilde{Q}_{\nu}-\tilde{p}_{1\nu}\widetilde{Q}_{\mu}),
\nn
\ea
where we introduce the following notations
\ba
\widetilde{Q}_{\mu}&=&Q_{\mu\nu}q_{\nu}-\frac{q_{\mu}}{q^2}\bar
{Q}, \ \ \widetilde{Q}_{\mu}q_{\mu}=0, 
\nn\\
\widetilde{Q}_{\mu\nu}&=& Q_{\mu\nu}+\frac{q_{\mu}q_{\nu}}{q^4}\bar Q-
\frac{q_{\nu}q_{\alpha}}{q^2}Q_{\mu\alpha}-
\frac{q_{\mu}q_{\alpha}}{q^2}Q_{\nu\alpha},
\label{eq:40}\\
\widetilde{Q}_{\mu\nu}q_{\nu} &=& 0, \ \bar
Q=Q_{\alpha\beta}q_{\alpha}q_{\beta}.\nn
\ea
The structure functions $V_i(Q^2) \ (i=1-5)$, which describe the
part of the hadronic tensor due to the tensor polarization of the
deuteron target, have the following form in terms of the deuteron
form factors
\ba
V_1(Q^2)&=&-G_M^2, \ \ V_5(Q^2)=0, \nn\\
V_2(Q^2)&=&G_M^2+\frac{4}{1+\tau }(G_C+\frac{\tau }{3}G_Q +\tau
G_M)G_Q, 
\label{eq:41}\\
V_3(Q^2)&=&-2\tau [G_M^2+2G_QG_M], \ \ V_4(Q^2)=4M^2\tau (1+\tau
)G_M^2.\nn
\ea
The fifth structure function $V_5(Q^2)$ is zero since deuteron form
factors are real functions in the considered kinematical region. In
the time-like region of momentum transfers this structure function
is not zero and it is given by $V_5(Q^2)=-4\tau
ImG_QG_M^*$.

The differential cross section of the reaction (\ref{eq:1}) describing the
scattering of unpolarized lepton beam on the tensor polarized deuteron
target can be written as
\be\label{eq:42}
\frac{d\sigma(Q)}{d\Omega}=\frac{d\sigma_{un}}{d\Omega}\left [1+A_{xx}(Q_{xx}-Q_{yy})+
A_{xz}Q_{xz}+A_{zz}Q_{zz}\right ],
\ee
where $A_{ij}$ are the asymmetries caused by the tensor polarization
of the deuteron target. Here we used the conditions that the tensor
$Q_{ij}$ is symmetrical and traceless: $Q_{xx}+Q_{yy}+Q_{zz}=0$. The
asymmetries have the following form in terms of the deuteron
electromagnetic form factors
\ba
DA_{xx}&=&\frac{1}{2}\frac{\vec{k}_2^2}{M^2}\frac{\sin^2\theta}{z}
\Bigl \{\left (\varepsilon^2_1+\tau M^2-m^2\right )G^2_M+4(1+\tau )^{-1}G_Q\Bigl [ \tau (M+\varepsilon_1)
\nn\\
&&
(\varepsilon_1-\tau M)G_M+ 
\left(\varepsilon^2_1-\tau M^2-2\tau
M\varepsilon_1\right )\left (G_C+\frac{\tau}{3}G_Q\right)\Bigr]\Bigr\}, 
\nn\\
DA_{xz}&=&-2\frac{\varepsilon_1|\vec{k}_2|}{M^2z}\sin\theta 
\Bigl \{ \Bigl [
\varepsilon_1 \left(|\vec{k}_1|-|\vec{k}_2|cos\theta \right )
\left(1+\tau\frac{M^2}{\varepsilon^2_1}-\frac{m^2}{\varepsilon^2_1}\right )-
\nn\\
&&
2\tau M|\vec{k}_1|\left(1+\frac{M}{\varepsilon_1}\right)\Bigr ]G^2_M
+4\frac{\varepsilon_1G_Q}{1+\tau}\Bigl [\left (|\vec{k}_1|-|\vec{k}_2|\cos\theta \right )
\nn\\
&&
\left(1-\tau \frac{M^2}{\varepsilon^2_1}
-2\tau \frac{M}{\varepsilon_1}\right )\left(G_C+\frac{\tau}{3}G_Q\right)+\tau \left(1-\tau \frac{M}{\varepsilon_1}\right )G_M 
\label{eq:43}
\\
&&
\left (|\vec{k}_1|\left (1-\tau
\frac{M}{\varepsilon_1}\right)-|\vec{k}_2|\left (1+\frac{M}{\varepsilon_1}\right )\cos\theta
\right) 
\Bigr ] \Bigr \},
\nn\\
DA_{zz}&=&\frac{1}{zM^2}\Bigl  \{
\left [\left (|\vec{k}_1|-|\vec{k}_2|\cos\theta \right )^2-\frac{1}{2}
\vec{k}_2^2\sin^2\theta \right ]
\Bigr [
\left (\varepsilon^2_1+\tau M^2-m^2\right )G^2_M
\nn\\
&&
+4(1+\tau )^{-1}G_Q \Bigl (\tau (M+\varepsilon_1)(\varepsilon_1-\tau M)G_M+
\nn\\
&&
\left (\varepsilon^2_1-\tau M^2-2\tau M\varepsilon_1\right ) \left(G_C+\frac{\tau}{3}G_Q\right ) \Bigr ) \Bigr]
+4\tau M|\vec{k}_1| G_M \Bigl [ M|\vec{k}_1|\
\nn\\
&&
(1+\tau )G_M-\left (|\vec{k}_1|-|\vec{k}_2|\cos\theta \right )
\Bigl( (M+\varepsilon_1)G_M+2(\varepsilon_1-\tau M) G_Q\Bigr )\Bigr ]\Bigr \}. \nn
\ea
In the limit of zero lepton mass we have the following expressions for
the asymmetries due to the tensor polarization of the deuteron
target
\ba
\bar D A^{(0)}_{xx}&=&\frac{\tau}{2}\Bigl \{
\left (1+\tau \frac{M^2}{\varepsilon^2_1}\right )G^2_M
+4(1+\tau )^{-1}G_Q\Bigl[\tau \left(1+\frac{M}{\varepsilon_1}\right)\left(1-\tau
\frac{M}{\varepsilon_1}\right)G_M 
\nn\\
&&
+\left(1-\tau \frac{M^2}{\varepsilon^2_1}-2\tau
\frac{M}{\varepsilon_1}\right)\left(G_C+\frac{\tau}{3}G_Q\right)\Bigr ]\Bigr \}, \nn\\
\bar DA^{(0)}_{xz}&=&-\frac{\varepsilon_2}{M}\frac{\tau}{1+\tau}\sin\theta
\Bigr \{
4\left(1+\frac{M}{\varepsilon_1}\right)G_Q\left(G_C+\frac{\tau}{3}G_Q\right)+
\nn\\
&&
(1+\tau)\left(1+\frac{M}{\varepsilon_1}\right)\tan^2\frac{\theta}{2}G^2_M+
2\left (1-\tau \frac{M}{\varepsilon_1}\right)\Bigr [-1-\tau
\label{eq:44}
\\
&&
+2\sin^2\frac{\theta}{2}\left (1+\frac{\varepsilon_1}{M}+\frac{\varepsilon^2_1}{M^2}-\tau
\frac{\varepsilon_1}{M}\right )\Bigr ]
\left(1+\tan^2\frac{\theta}{2}\right )G_MG_Q
\Bigr \},
\nn\\
\bar DA^{(0)}_{zz}&=&-\frac{\tau}{2}
\left \{
\left [ 6\frac{\tau}{1+\tau}\frac{\varepsilon_1+\varepsilon_2}{\varepsilon_1}\left (1+\frac{M}{\varepsilon_1}\right )G_Q-G_M
\right ]G_M+ \right .
\nn\\
&&
+\tan^2\frac{\theta}{2}\left [1-2\tau -6\tau
\frac{M}{\varepsilon_1}\left (1+\frac{M}{2\varepsilon_1}\right )\right ]
\nn\\
&&
\left . \left [G^2_M+\frac{4}{1+\tau}\cot^2\frac{\theta}{2}G_Q\left (G_C+\frac{\tau}{3}G_Q\right) \right ]
\right  \}. 
\nn
\ea
In the coordinate system where the
$Z$ axis is directed along the virtual photon momentum, the asymmetries due to the tensor polarization of the deuteron target have the following form:
\be\label{eq:45}
A_{\alpha}=T_{\alpha\beta}(\psi)A_{\beta},
\ee
where the indices of the rotation matrix have following meaning:
$\alpha =ZZ$, $XX$, $XZ$ and $\beta =zz$, $xx$, $xz$. The rotation matrix
can be written as
\be\label{eq:46}
T(\psi)=\left(\begin{array}{ccc}\frac{1}{4}(1+3\cos{2\psi})&\frac{3}{4}(1-\cos{2\psi})
&\frac{3}{4}\sin{2\psi}\\\frac{1}{4}(1-\cos{2\psi})&\frac{1}{4}(3+\cos{2\psi})&
\frac{-1}{4}\sin{2\psi}\\-\sin{2\psi}&\sin{2\psi}&\cos{2\psi}\end{array}\right)
\ee

Along with the transformation of the asymmetries it is necessary to
transform the tensor of the quadrupole polarization which describes
the tensor polarization of the deuteron target. The new tensor
polarization parameters are related to $Q_{zz}$,
$(Q_{xx}-Q_{yy})$, and $Q_{xz}$,  Eq. (\ref{eq:18}), as follows:
\ba
Q_{ZZ}&=&\frac{1}{4}(1+3\cos2\psi )Q_{zz}+\frac{1}{4}(1-\cos2\psi
)(Q_{xx}-Q_{yy})+\sin2\psi Q_{xz}, 
\nn\\
Q_{XX}-Q_{YY}&=&\frac{3}{4}(1-\cos2\psi
)Q_{zz}+\frac{1}{4}(3+\cos2\psi )(Q_{xx}-Q_{yy})+\sin2\psi Q_{xz},
\nn
\\
Q_{XZ}&=&-\frac{3}{4}\sin2\psi Q_{zz}+\frac{1}{4}\sin2\psi
(Q_{xx}-Q_{yy})+\cos2\psi Q_{xz}.
\label{eq:47}
\ea
\section{Numerical estimation}
In this section we give numerical estimations of the effect of the mass on the kinematical variables and on some of the experimental observables. As two variables define completely the kinematics for a binary process, the results are preferentially illustrated as bi-dimensional plots as function of the muon beam energy and the muon scattering angle.
 
\subsection{Kinematics} 

The effect of the lepton mass on the kinematical variables is illustrated for the scattered lepton energy and for the momentum transfer squared.

\begin{figure}
    \centering
   \mbox{
        {\includegraphics[width=0.4\textwidth]{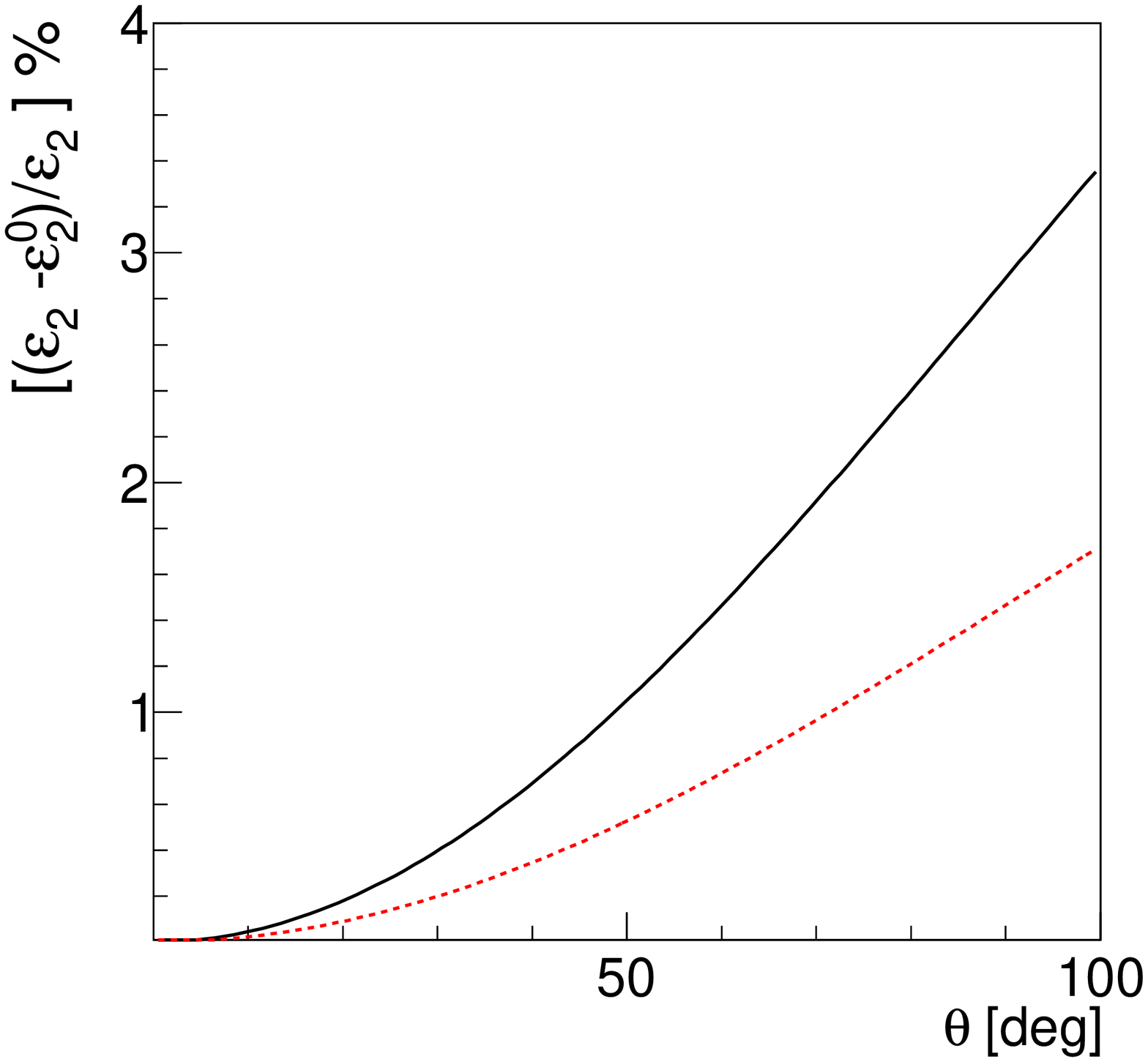}
\label{Fig.fig1}}
        \quad\quad
        {\includegraphics[width=0.4\textwidth]{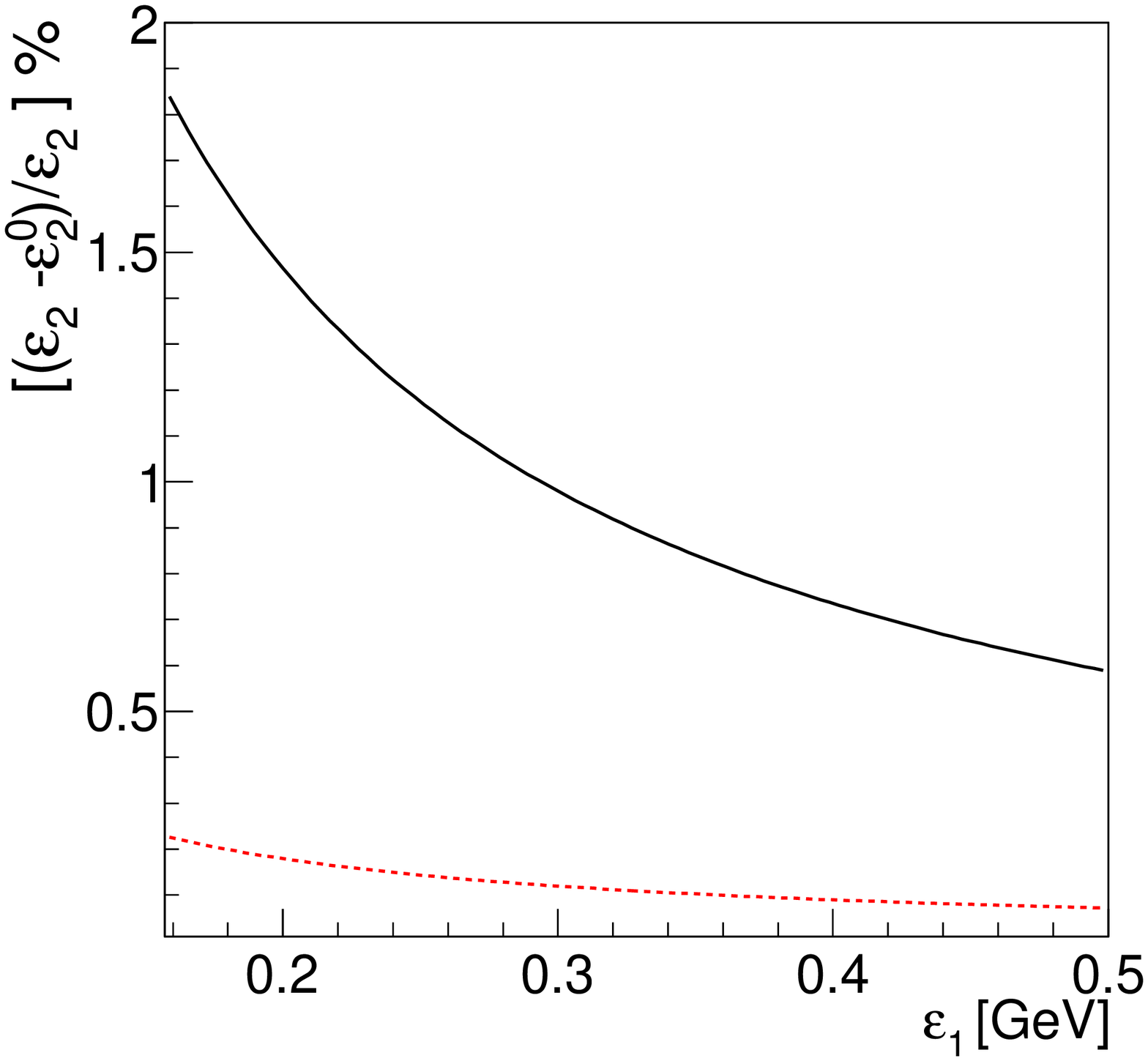}
\label{Fig.fig2}}
   }
    \caption{Effect of the lepton mass on the kinematics of the scattered lepton. The relative difference (in \%) between the scattered lepton energy taking and not taking into account the lepton mass is shown as a function of the scattering muon angle (left) for two values of the incident energy: $\epsilon_1= 50$ MeV (solid black line) and $\epsilon_1= 200$ MeV (dashed red line) and as function of the incident energy (right) for two values of the lepton scattering angle $\theta=60^\circ$, (solid black line) and 
$\theta=20^\circ$, (dashed red line). }
    \label{fig.Kinematics}
\end{figure}

The relative difference between the scattered lepton energy taking and not taking into account the lepton mass is shown in Figs. \ref{fig.Kinematics} as a function of the scattering muon angle for two values of the incident energy: $\epsilon_1= 50$ MeV (solid black line), and $\epsilon_1= 200$ MeV (dashed red line), and as function of the incident energy for two values of the lepton scattering angle $\theta=60^\circ$ (solid black line), and 
$\theta=20^\circ$ (dashed red line). 

\begin{figure}
    \centering
    \mbox{
        {\includegraphics[width=0.8\textwidth]{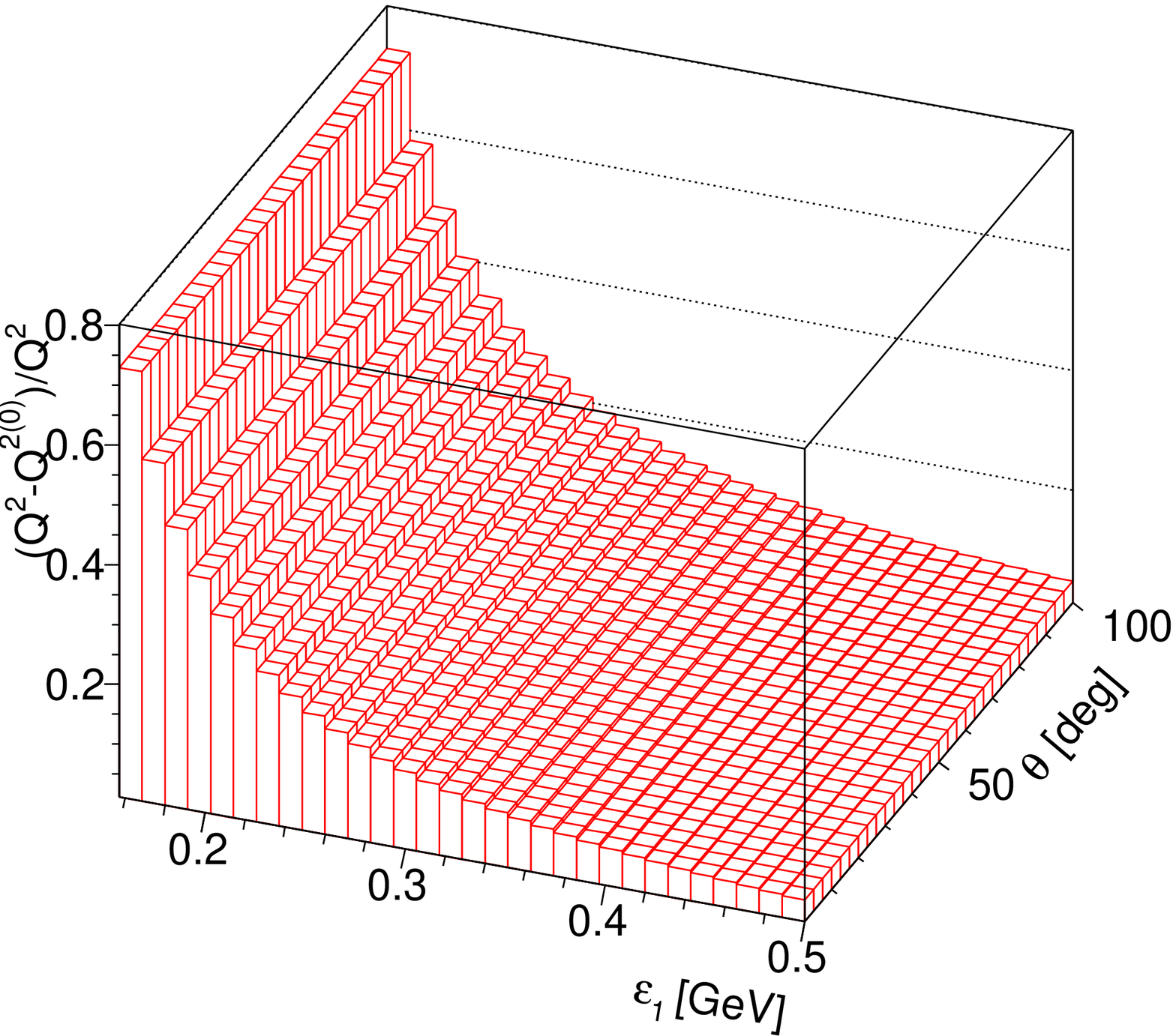}}
   }
    \caption{Bi-dimensional plot of the relative difference of momentum transfer squared as function of the incident energy and of the muon scattering angle.}
    \label{Fig.Q2}
\end{figure}

The effect of the mass on the momentum transfer squared is shown in Fig. \ref{Fig.Q2} as a bi-dimensional plot as a function of the incident energy and on the muon scattering angle.

From these figures it appears that, the effect of the lepton mass  on the energy and angle of the scattered muon, as well as on the momentum transfer squared, is  sizable in the considered kinematical range, in particular at low beam energies.

\subsection{Parametrization of the deuteron form factors}
For the calculation of unpolarized and polarized observables the knowledge of the deuteron form factors is needed. We used the parametrization from Ref. \cite{TomasiGustafsson:2005ni}  which is based on a "two--component model" of the deuteron, inspired from vector meson dominance \cite{IJL73} where the $pn$ core is surrounded by an (isoscalar) meson cloud. This parametrization  has a simple analytical form, and reproduces at best the existing experimental data. We recall here the formulas and the parameter set that we used.
 
The three deuteron form factors are parametrized as: 
\be
G_i(Q^2)=N_i g_i(Q^2) F_i(Q^2),~i=C,Q,M, 
\label{eq:eq1}
\ee
where $N_i$ is the normalization of the $i$-th form factor at $Q^2=0$:
$$N_C=G_C(0)=1,~N_Q=G_Q(0)=M^2{\cal Q}_d=25.83,~
N_M=G_M(0)=\displaystyle\frac{M}{m_p}\mu_d=1.714.$$
Here ${\cal Q}_d$, and $\mu_d$ are the quadrupole and the magnetic moments of the deuteron, $m_p$ is the proton mass.

The expression for the meson cloud is:  
$$F_i(Q^2)= 1-\alpha_i-\beta_i+
\alpha_i\displaystyle\frac{m_{\omega}^2}{m_{\omega}^2+Q^2}
+\beta_i\displaystyle\frac{m_{\phi}^2}{m_{\phi}^2+Q^2}, $$
where $m_{\omega}=0.784$ GeV ($m_{\phi}=1.019$ GeV) is the mass of the $\omega$ ($\phi$)-meson. These expressions are built in such form that $F_i(0)=1$, for any values of the free parameters $\alpha_i$ and $\beta_i$, which are real numbers.

The intrinsic core is parametrized as: 
\begin{equation}
 g_i(Q^2)=1/\left [1+\gamma_i{Q^2}\right ]^{\delta_i}.
\label{eq:eq12}
\end{equation} 
The terms $g_i(Q^2)$ are  functions of two  parameters, also real. We took common values for all form factors: $\gamma_i=12.1$ and $\delta_i=1.05$.

The parameters $\alpha_C$ and $\alpha_M$ have been fixed by the value of the experimental node which appears at $Q^2=1.9$ GeV$^2$ and $\simeq 0.7$ Ge$V^2$ for $G_M$ and $G_C$ respectively. The other parameters are: $\beta_i(G_C)=-5.11$, $\alpha_Q=4.21$, $\beta_Q$=-3.41, and $\beta_M$=-2.86. 

\subsection{Experimental observables}

\begin{figure}
    \centering
    \mbox{
        {\includegraphics[width=0.8\textwidth]{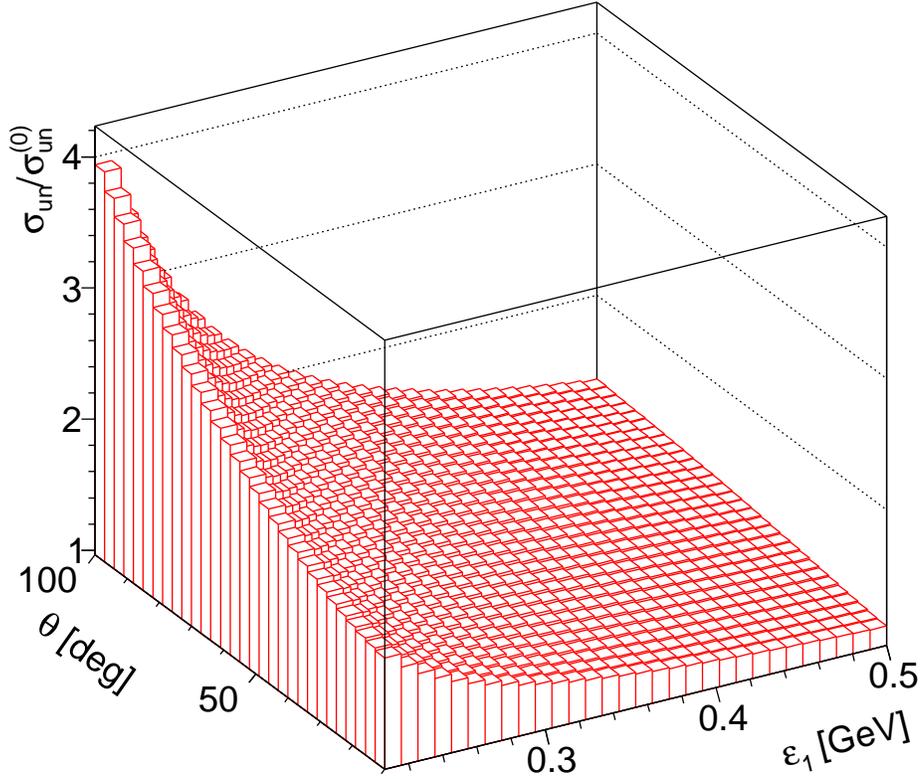}}
   }
    \caption{Bi-dimensional plot of the ratio between the cross section taking and not taking into account the lepton mass is shown as function of $\varepsilon_1$ and $\theta$}
    \label{Fig:sigma}
\end{figure}

Using Eqs. (23-29) with the parametrization of form factors above described, the unpolarized cross section is calculated in the relevant kinematical range. The correction to the Born cross section due to the finite lepton mass is illustrated in Fig. \ref{Fig:sigma}, where the ratio between the cross section taking and not taking into account the lepton mass is shown as function of $\varepsilon_1$ and $\theta$. One can see that this ratio increases essentially at small energies and large angles.

\begin{figure}
    \centering
    \mbox{
        {\includegraphics[width=0.8\textwidth]{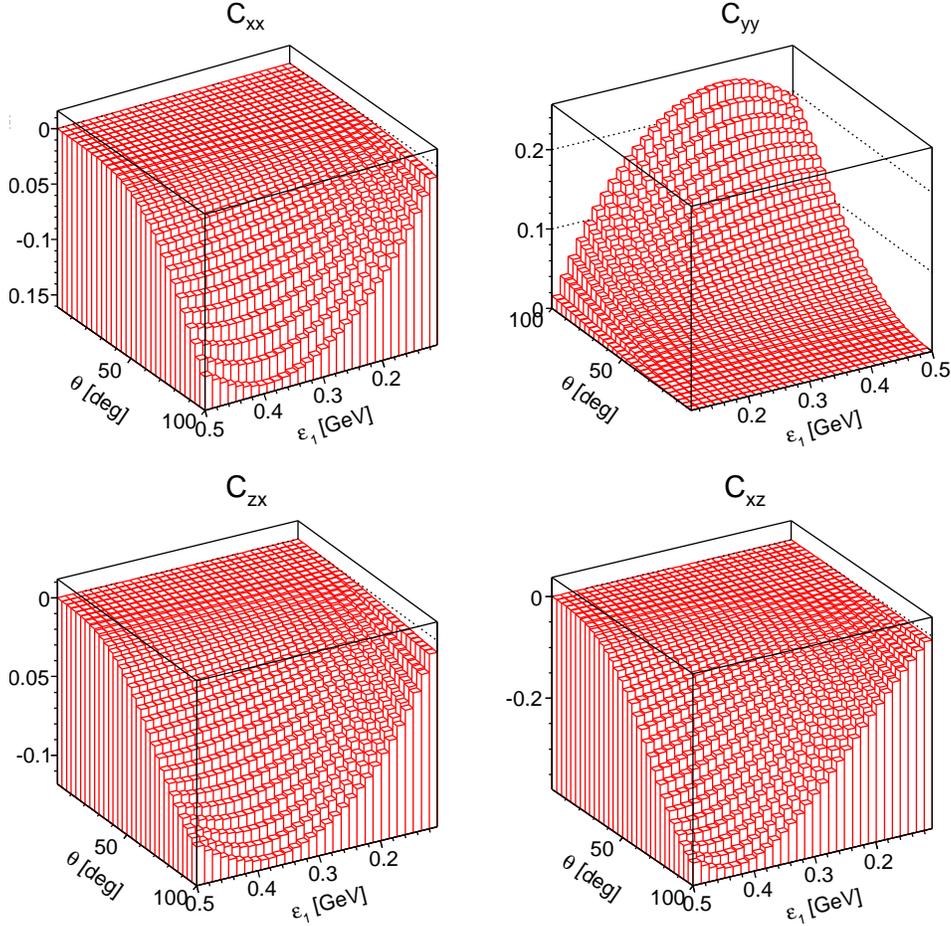}}
   }
    \caption{Correlation coefficients $C_{xx}$ (top left), $C_{yy}$, (top right) $C_{zx}$, (bottom left), and $C_{xz}$ (bottom right) as functions of  $\varepsilon_1$ and $\theta$. }
    \label{Fig:corr1}
\end{figure}

The polarization observables $C_{xx}$, $C_{yy}$, $C_{zx}$, $C_{xz}$,  induced by a polarized lepton beam on a vector polarized deuteron target are illustrated in Fig. \ref{Fig:corr1}, as bi-dimensional plots in function of $\varepsilon_1$ and $\theta$.
 The correlation coefficients vanish for $\theta=0$ and at small energies. $C_{xx}$, $C_{zx}$, and $C_{xz}$  become sizable and negative as the angle and energy increase, whereas  $C_{yy}$ becomes positive.

\begin{figure}
    \centering
    \mbox{
        {\includegraphics[width=0.8\textwidth]{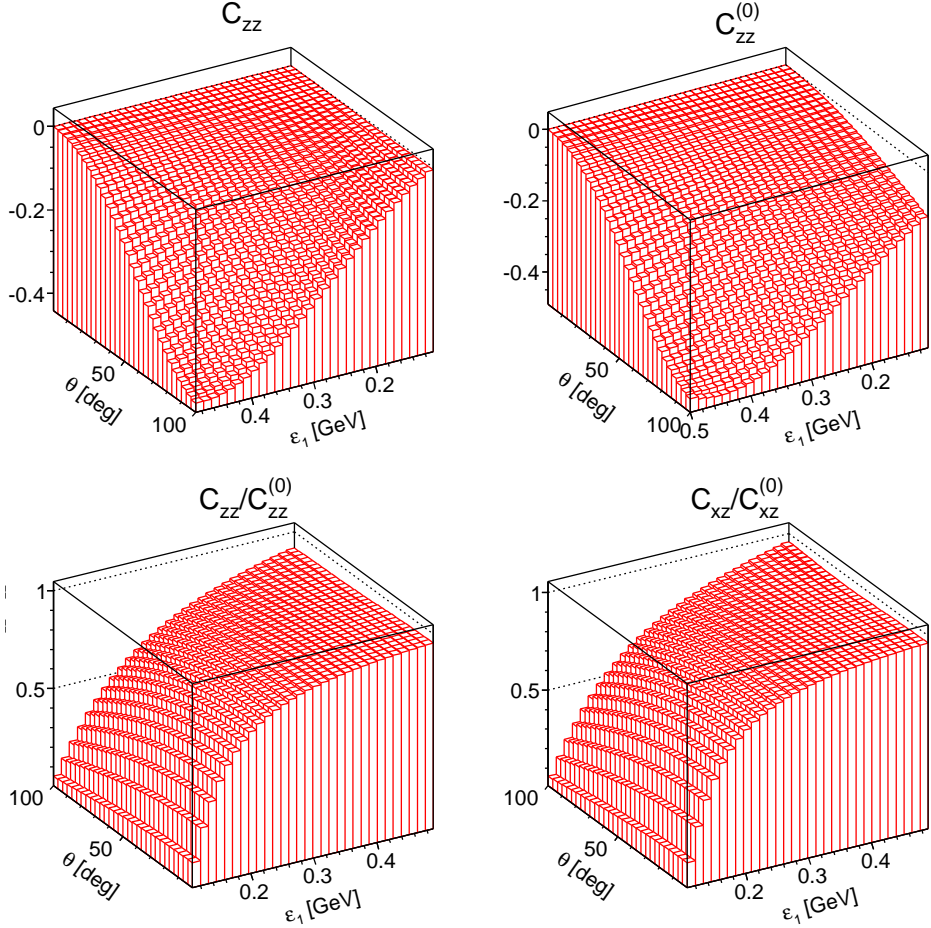}}
   }
    \caption{Correlation coefficients $C_{zz}$ (top left), $C_{zz}^{(0)}$, (top right) as function of $\varepsilon_1$ and $\theta$. The ratio of the observables taking and not taking into account the lepton mass is shown for $C_{zz}$, (bottom left), and $C_{xz}$ (bottom right). }
    \label{Fig:corr2}
\end{figure}

In Fig. \ref{Fig:corr2}, the polarization coefficient $C_{zz}$ is shown (top left). The same coefficient setting the mass of the lepton to zero is shown (top right). The ratio of the observables taking and not taking into account the lepton mass is shown for $C_{zz}$, (bottom left), and $C_{xz}$ (bottom right). Again, the relative effect of the mass is very large at low energies and large angles.

\begin{figure}
    \centering
    \mbox{
        {\includegraphics[width=0.99\textwidth]{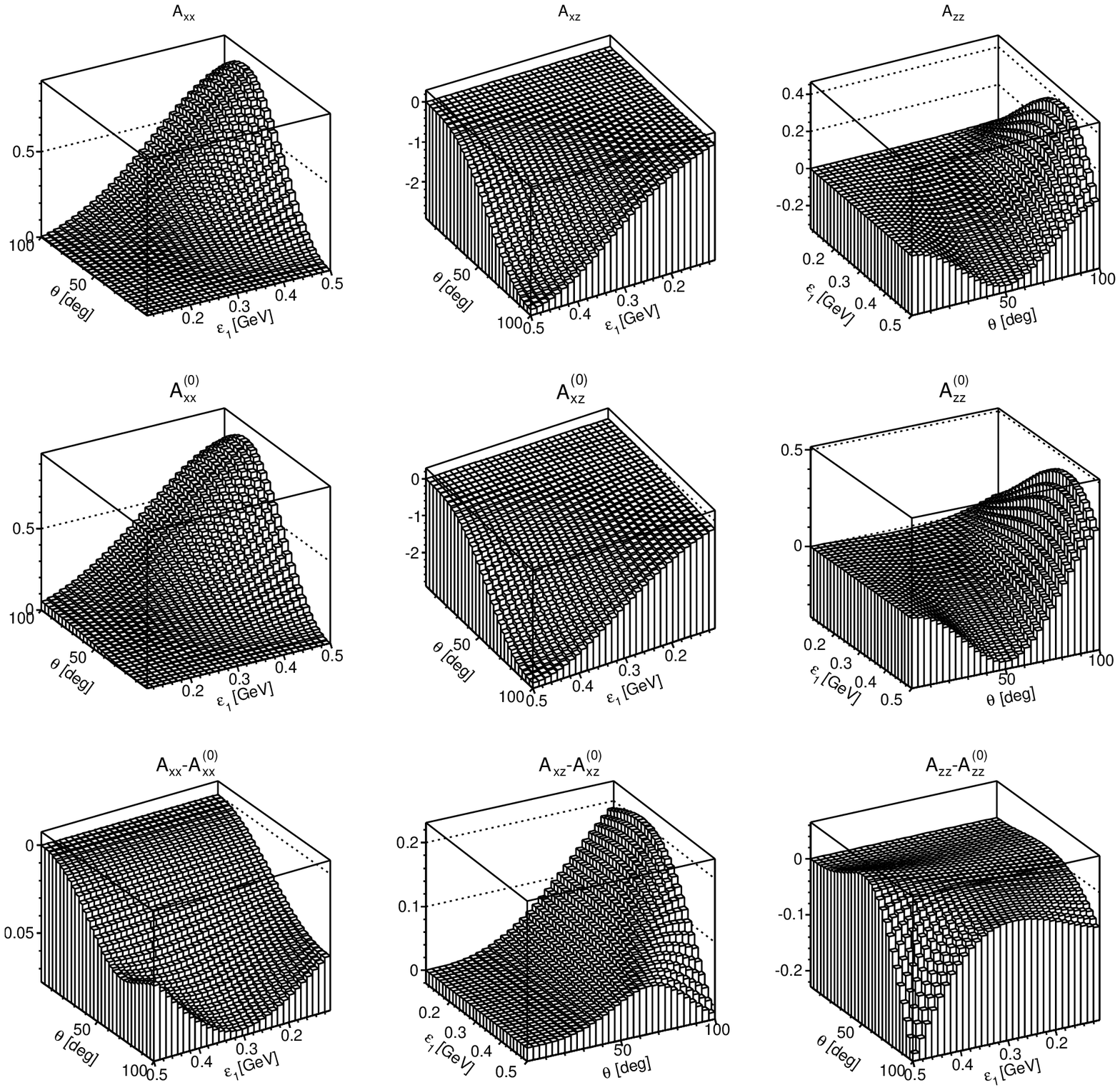}}
   }
    \caption{Tensor asymmetries from left to right $A_{xx}$, $A_{xz}$, $A_{zz}$, are shown. From top to bottom, the first(second) row represents the observables taking (not taking) into account the lepton mass. The third row shows their difference. }
    \label{Fig:tensor}
\end{figure}

The tensor asymmetries induced by a tensor polarized deuteron target and unpolarized lepton beam are illustrated in 
Fig. \ref{Fig:tensor}. From left to right $A_{xx}$, $A_{xz}$, and $A_{zz}$ are shown. From top to bottom, the first(second) row represents the observables taking (not taking) into account the lepton mass. The third row shows their difference. We do not plot the ratio, as it diverges for the observable $A_{zz}$, which  changes of sign in the considered range. The relative effect of the mass is of about 10\% on $A_{xx}$ $A_{xz}$ and can reach 50\% on 
$A_{zz}$. 

\section{Conclusion}

We have calculated the polarized and unpolarized cross section for lepton deuteron elastic scattering, taking into account the lepton mass and applying to the case of muon scattering.

Besides the unpolarized cross section, different observables have been calculated, according to the possible polarization of the lepton beam and the deuteron target.  
The spin correlation coefficients due to the
lepton beam polarization and to the vector polarization of the deuteron
target, as well as the asymmetries due to the tensor polarization of the
deuteron target have been explicitly derived  The calculations have been done for two coordinate systems: in
the first one the $z$ axis is directed along the lepton beam momentum and, in the second one, along the virtual photon momentum.
	
The numerical application needs a parametrization of the deuteron form factors. We chose the parametrization from Ref. \cite{TomasiGustafsson:2005ni} which is based on a two component model of the deuteron, where the $pn$ core is surrounded by an (isoscalar) meson cloud. This model reproduces very well the existing experimental data.

It is shown that the effect of the finite lepton mass is sizable in particular at low incident energies and large scattering angles.

These results are particularly important in relation to planned measurements of low energy muon-deuteron scattering, which aim to a precise determination of the charge radius.

\section{Appendix}

\setcounter{equation}{0}
\def\theequation{A.\arabic{equation}}

We give here the expressions which relate the description of the
polarization state of the deuteron target for different approaches.
For the case of arbitrary polarization of the target, the deuteron polarization is described by the spin-density matrix which is defined, in the general case, by 8 parameters.

\subsection{Coordinate representation}.

The deuteron spin-density matrix in the coordinate representation
has the form
\ba
\rho_{\mu\nu}&=&-\frac{1}{3}\left (g_{\mu\nu}-\frac{p_{\mu}p_{\nu}}{M^2}\right )
+\frac{i}{2M}\varepsilon_{\mu\nu\lambda\rho} s_{\lambda}p_{\rho}+
Q_{\mu\nu},~ Q_{\mu\nu}=Q_{\nu\mu},~Q_{\mu\mu}=0, \nn\\
&&~p_{\mu}Q_{\mu\nu}=0, 
\label{eq:A1}
\ea
where $p_{\mu }$ is the deuteron
four-momentum, $s_{\mu}$ and $Q_{\mu\nu}$ are the deuteron
polarization four-vector describing the vector polarization and the
deuteron quadrupole-polarization tensor describing the tensor
polarization.

In the deuteron rest frame Eq. (\ref{eq:A1}) becomes
\be\label{eq:A2}
\rho_{ij}=\frac{1}{3}\delta_{ij}-\frac{i}{2}\varepsilon
_{ijk}s_k+Q_{ij}, \ ij=x,y,z.
\ee

\subsection{Helicity representation}

The spin-density matrix can be written in the helicity
representation using the following relation
\be\label{eq:A3}
\rho_{\lambda\lambda'}=\rho_{ij}e_i^{(\lambda )*}e_j^{(\lambda')}, \
\rho_{\lambda\lambda'}= (\rho_{\lambda'\lambda})^* \,, \lambda
,\lambda'=+,-,0,
\ee
where $e_i^{(\lambda )}$ are the deuteron spin functions which have
the deuteron spin projection $\lambda $ onto the quantization axis
($z$ axis). They are
\be\label{eq:A4}
e^{(\pm )}=\mp \frac{1}{\sqrt{2}}(1,\pm i,0), \ e^{(0)}=(0,0,1).
\ee
The elements of the spin-density matrix in the helicity
representation are related to the ones in the coordinate
representation as follows
\ba
\rho _{++}&=&\frac{1}{3}+ \frac{1}{2}s_z-\frac{1}{2}Q_{zz}, 
\rho_{--}=\frac{1}{3}- \frac{1}{2}s_z-\frac{1}{2}Q_{zz}, ~\rho_{00}=\frac{1}{3}+Q_{zz},
\nn\\
\rho_{+-}&=&-\frac{1}{2}(Q_{xx}-Q_{yy})+iQ_{xy},~
\rho_{+0}=\frac{1}{2\sqrt{2}}(s_x-is_y)-
\frac{1}{\sqrt{2}}(Q_{xz}-iQ_{yz}),
\nn\\
\rho_{-0}&=&\frac{1}{2\sqrt{2}}(s_x+is_y)-
\frac{1}{\sqrt{2}}(Q_{xz}+iQ_{yz})\,.
\label{eq:A5}
\ea
To obtain this relations we used the condition
$Q_{xx}+Q_{yy}+Q_{zz}=0\,.$

\subsection{Representation in terms of the population numbers}

The description of the polarized deuteron target in terms of the population
numbers $n_+$, $n_-$, and $n_0 $ is often used in the formulation of spin
experiments (see, for example, Ref. \cite{A05}). Here $n_+, \ $ $n_-
$ and $n_0 $ are the fractions of the atoms in the polarized target
with the nuclear spin projection on to the quantization axis $m=+1,
\ $ $m=-1$ and $m=0,$ respectively. If the spin-density matrix is
normalized to 1, i.e., $Tr~\rho =1$, then we have $n_++n_-+n_0=1.$
Thus, the polarization state of the deuteron target is defined in
this case by two parameters: the so-called V (vector) and T (tensor)
polarizations
\be\label{eq:A6}
V=n_+-n_-, \ T=1-3n_0.
\ee
Using the definitions for the quantities $n_{\pm ,0}$
\be\label{eq:A7}
n_{\pm }=\rho_{ij}e_i^{(\pm )*}e_j^{(\pm )}, \
n_0=\rho_{ij}e_i^{(0)*}e_j^{(0)},
\ee
we have the following relation between $V$ and $T$ parameters and
parameters of the spin-density matrix in the coordinate
representation (in the case when the quantization axis is directed
along the z axis)
\be\label{eq:A8}
n_0=\frac{1}{3}+Q_{zz}, \ n_{\pm }=\frac{1}{3}\pm \frac{1}{2}s_z-
\frac{1}{2}Q_{zz},
\ee
or
\be\label{eq:B9}
T=-3Q_{zz}, \ V=s_z.
\ee

\subsection{Representation of the spherical tensors}

Let us relate now the parameters of the spin-density matrix in the
coordinate representation to the parameters of the matrix in the
representation of the spherical tensors.

According to the Madison Convention \cite{MC70}, the spin-density matrix
of a spin-one particle is given by the expression
\be\label{eq:A10}
\rho =\frac{1}{3}\sum_{kq}t^*_{kq}\tau_{kq},
\ee
where $t_{kq}$ are the polarization parameters of the deuteron
spin-density matrix and $\tau_{kq}$ are the spherical tensors. The
spherical tensors are expressed as
$$
\tau_{00}=1, \ \tau_{10}=\sqrt{\frac{3}{2}}S_z, \ \tau_{1\pm
1}=\mp \frac{\sqrt{3}}{2}(S_x\pm iS_y),~
\tau_{20}=\frac{3}{\sqrt{2}}(S^2_z-\frac{2}{3}),
$$
\be
\tau_{2\pm 2}=\frac{\sqrt{3}}{2}(S_x\pm iS_y)^2,~
\tau_{2\pm 1}=\mp \frac{\sqrt{3}}{2}[(S_x\pm iS_y)S_z+S_z(S_x\pm
iS_y)],
\label{eq:A11}
\ee

\be\label{eq:A12}
S_x=\frac{1}{\sqrt{2}} \left(\begin{array}{ccc}
0&1&0\\
1&0&1\\
0&1&0
\end{array}\right), \ S_y=\frac{1}{\sqrt{2}} \left(\begin{array}{ccc}
0&-i&0\\
i&0&-i\\
0&i&0
\end{array}\right), \ S_z=\left(\begin{array}{ccc}
1&0&0\\
0&0&0\\
0&0&-1
\end{array}\right).
\ee

From Eq. (A.11) and the hermiticity of the spin operator it is straightforward to get
\be\label{eq:A13}
\tau^+_{kq}=(-1)^q\tau_{k-q}.
\ee
The hermiticity condition for the density matrix yields for
$t_{kq}$:
\be\label{eq:A14}
t^*_{kq}=(-1)^qt_{k-q}.
\ee
From this equation one can see that
\be
t^*_{10}=t_{10}, \  t^*_{11}=-t_{1-1}, \ t^*_{20}=t_{20}\,,~
t^*_{22}=t_{2-2}, \ t^*_{21}=-t_{2-1},
\label{eq:A15}
\ee
i.e., the parameters $t_{10}$ and $t_{20}$ are real ones, and the
parameters $t_{11}$, $t_{21}$ and $t_{22}$ are complex ones. So, in
total there are 8 independent real parameters as it is required for spin-one massive particles.

The explicit expression of the deuteron density
matrix is:
\be\label{eq:A16}
\rho =\frac{1}{3}\left(\begin{array}{ccc}
1+\sqrt{\frac{3}{2}}t_{10}+\frac{1}{\sqrt{2}}t_{20}
&\sqrt{\frac{3}{2}}(t_{1-1}+t_{2-1})&\sqrt{3}t_{2-2}\\
-\sqrt{\frac{3}{2}}(t_{11}+t_{21})&1-\sqrt{2}t_{20}&
\sqrt{\frac{3}{2}}(t_{1-1}-t_{2-1})\\
\sqrt{3}t_{22}&-\sqrt{\frac{3}{2}}(t_{11}-t_{21})&
1-\sqrt{\frac{3}{2}}t_{10}+\frac{1}{\sqrt{2}}t_{20}
\end{array}\right).
\ee
The density matrix is normalized to 1, i.e., $Tr~\rho =1$. Using the
expression for the density matrix in the helicity representation,
Eq. (\ref{eq:A5}), we get the following relations between the parameters of
the density matrix in the coordinate representation and spherical
tensor one

\ba
t_{10}&=&\sqrt{\frac{3}{2}}s_z,~
Re~t_{11}=-Re~ t_{1-1}=-\frac{\sqrt{3}}{2}s_x,~
\nn\\
Im~t_{11}&=&Im ~t_{1-1}=-\frac{\sqrt{3}}{2}s_y,
t_{20}=-\frac{3}{\sqrt{2}}Q_{zz},
\nn\\
Re~t_{21}&=&-Re~t_{2-1}=\sqrt{3}Q_{xz},~
Im~t_{21}=Imt_{2-1}=\sqrt{3}Q_{yz}, 
\label{eq:A17}\\
Re~t_{22}&=&Re~t_{2-2}=-\frac{\sqrt{3}}{2}(Q_{xx}-Q_{yy}),~
Im~t_{22}=-Im~t_{2-2}=-\sqrt{3}Q_{xy}. 
\nn
\ea

\end{document}